\documentclass[fleqn,twoside]{article}
\usepackage{espcrc2}

\usepackage{epsfig}
\usepackage{graphicx}
\usepackage[figuresright]{rotating}

\newcommand{\be}{\begin{equation}}
\newcommand{\ee}{\end{equation}}
\newcommand{\ba}{\begin{eqnarray}}
\newcommand{\ea}{\end{eqnarray}}

\newcommand{\AmS}{{\protect\the\textfont2
  A\kern-.1667em\lower.5ex\hbox{M}\kern-.125emS}}
\def\spose#1{\hbox to 0pt{#1\hss}}
\def\ltapprox{\mathrel{\spose{\lower 3pt\hbox{$\mathchar"218$}}
 \raise 2.0pt\hbox{$\mathchar"13C$}}}

\def\lsim{\raise0.3ex\hbox{$<$\kern-0.75em\raise-1.1ex\hbox{$\sim$}}}
\def\gsim{\raise0.3ex\hbox{$>$\kern-0.75em\raise-1.1ex\hbox{$\sim$}}}

\title{A Geometrical Interpretation of
              Hyperscaling Breaking in the Ising Model
       \thanks{This work was partially supported by DFG Grant No. FOR 339/1-2.}}

\author{G. Andronico\address{Dipartimento di Fisica e Astronomia, 
Universita' di Catania \& INFN, I-95129 Catania, Italy},
 A. Coniglio\address{Universita' di Napoli "Federico II" \&
 INFM, Monte S. Angelo - Via Cintia, 80126 Napoli, Italy }, 
S. Fortunato\address{Fakult\"at f\"ur Physik, Universit\"at Bielefeld, D-33615 Bielefeld,
 Germany}}

\begin{document}
\begin{abstract}

In random percolation one finds that the mean field regime
above the upper critical dimension 
can simply be explained through the coexistence of  
infinite percolating clusters at the critical point. Because of the mapping
between percolation and critical behaviour in the Ising model, one might 
check whether the breakdown of hyperscaling in the Ising model can 
also be intepreted as due to an infinite multiplicity of percolating 
Fortuin-Kasteleyn clusters at the
critical temperature $T_c$. Preliminary results 
suggest that the scenario is much more involved than expected
due to the fact that the percolation variables behave differently 
on the two sides of $T_c$.

\end{abstract}

\maketitle

\section{INTRODUCTION}

          The critical behaviour of statistical mechanical systems reduces
          to mean field theory above the upper critical dimension $d_u$.
          In this case, the scaling relations between the critical
          exponents of the phase transition remain valid
          except the so-called hyperscaling relations, i.e., the
          equalities in which the number $d$ of space dimensions of the system
          explicitly appears, like

          \begin{eqnarray}
            2-\alpha={\nu}d.
            \label{eq1}
          \end{eqnarray}

          For this reason one says that, above $d_u$, {\it hyperscaling breaks}. 
          Some time ago, Coniglio \cite{coni} proposed an
          interpretation of hyperscaling breaking for the pure random
          percolation problem: above the upper critical dimension 
          the number of percolating clusters at the critical threshold, which is
          finite for $d\,{\leq}\,d_u$, becomes infinite. The presence
          of infinitely many interpenetrating clusters damps
          the fluctuations of the order parameter establishing the mean field
          regime. 
          
          The magnetization transition of the Ising model can be equivalently described 
          as a percolation transition of suitably defined site-bond clusters 
          \cite{FK}. In particular, the magnetization coincides with the percolation
          order parameter. 
          That suggests an analogous interpretation of hyperscaling breaking
          as for random percolation: we tried to test this conjecture.
          We present here preliminary results, based 
          on Monte Carlo simulations, concerning the multiplicity of percolating
          clusters at criticality both in random percolation and in the Ising model above the
          upper critical dimension $d_u$.

\section{BREAKDOWN OF HYPERSCALING IN RANDOM PERCOLATION}

          We start from an hypercubic lattice with 
          a fraction $p$ of occupied sites. 
          Near the critical density $p_c$
          the singular part of the cluster number $K(p)|_{sing}$ behaves as

          \begin{eqnarray}
            K(p)|_{sing}=\sum_{s}{n(s,p)}|_{sing}\,\sim\,|p-p_c|^{2-\alpha},
            \label{eq2}
          \end{eqnarray}
          
          ($n(s,p)$ is the number of clusters of size $s$).
          
          The connectedness length $\xi(p)$, i.e. the typical
          radius of the largest finite cluster, diverges as

          \begin{eqnarray}
            \xi(p)\,\sim\,|p-p_c|^{-\nu}.
            \label{eq3}
          \end{eqnarray}
          
          From Eqs. (\ref{eq3}) and (\ref{eq2}) one obtains

          \begin{eqnarray}
            K(p)|_{sing}\,\sim\,\xi^{(\alpha-2)/\nu}.
            \label{eq4}
          \end{eqnarray}

          Let us now assume that the singular
          behaviour comes only from the critical clusters, i.e.
          the clusters whose radius is about $\xi$. 
          Say $N_{\xi}$ the
          number of such clusters in a volume of the order $\xi^{d}$. The
          singular part of the cluster number is given by
          
          \begin{eqnarray}
            \frac{N_{\xi}}{\xi^d}\,{\simeq}\,K(p)|_{sing}\,\sim\,\xi^{(\alpha-2)/\nu}.
            \label{eq5}
          \end{eqnarray}
 
          If $N_{\xi}$ is of the order of unity, from Eq. (5)
          we obtain the hyperscaling relation (1); if instead 
          $N_{\xi}$ grows with $\xi$, so that, at the critical
          point, $N_{\xi}\,\rightarrow\infty$,
          then hyperscaling breaks down.
           
          When there are infinitely many spanning clusters the 
          percolation order parameter fluctuates over a distance 
          $\xi_1\,\ll\,\xi$; this would simply explain why above 
          $d_u$ the mean field solution is valid. Besides,
          one finds that,
          above the upper critical dimension (for random percolation
          $d_u=6$),

          \begin{eqnarray}
            N_{\xi}\,{\sim}\,\xi^{d-d_u}=\xi^{d-6}.
            \label{eq6}
          \end{eqnarray}

\vspace*{0.1cm}
\section{BREAKDOWN OF HYPERSCALING IN THE ISING MODEL}

          If one builds clusters by joining nearest-neighbouring spins of the
          same sign with the bond probability 
          $p_B=1-\exp(-2J/kT)$ ($J$ is the Ising coupling), 
          the magnetization transition of the Ising model becomes
          equivalent to the percolation transition of such clusters
          (called Fortuin-Kasteleyn or FK clusters): the 
          percolation temperature coincides with the thermal critical point $T_c$
          and the critical percolation exponents are just the Ising exponents.
          Because of that one can interpret the Ising transition 
          as a simple geometrical phenomenon.
          Therefore, the argument used above for the pure percolation problem
          can be extended to the Ising case. 
          One expects then that the 
          number of FK clusters $N_{\xi}$ in a region of linear dimension
          $\xi$ is of the order of unity below $d_u=4$, while above
          $d_u$ 

          \begin{eqnarray}
            N_{\xi}\,\sim\,\xi^{d-4}.
            \label{eq7}
          \end{eqnarray}

          This implies that, above $d_u$, there should be an infinite
          number of percolating FK clusters at the critical temperature.
          
          We warn that some care needs to be taken. 
          The percolation transition of the FK clusters
          in the Ising model is asymmetric
          on the two sides of $T_c$: 
          for $T>T_c$, the percolation variables coincide with the 
          thermal variables in any dimension, whereas 
          for $d>d_u=4$ and $T<T_c$, 
          the thermal fluctuations are different from the cluster fluctuations.
          There are indeed arguments suggesting that 
          the critical percolation
          exponents below $T_c$, except $\beta$, are different from
          the Ising ones for $d>4$ and that the
          upper critical dimension of the FK clusters
          is $d^{FK}_u(T<T_c)=6$, like in random percolation.

          \section{RESULTS FOR RANDOM PERCOLATION}

          The first numerical studies on the multiplicity of 
          infinite clusters at and above the upper critical dimension
          were carried on by de Arcangelis \cite{dearc}, but only
          small lattices could be used (up to $6^7$). 

          We performed Monte Carlo simulations
          for pure random site percolation  
          in several dimensions on hypercubic
          lattices.
          In any space dimension we calculated the number of distinct
          spanning clusters for different lattices at the critical density $p_c$.
          We noticed that, for $d\,{\leq}\,6$, there is
          almost always just a single percolating 
          cluster (or none), independently of the lattice size, whereas
          for $d=7$ we observed
          an "explosion" of the multiplicity 
          of percolating clusters, which increases with the lattice side $L$. 
          Our aim is to prove Eq. (6) which, 
          at the critical point, assumes the following finite-size scaling form:

          \begin{eqnarray}
            N_{\infty}\,\propto\,L^{d-6}.
            \label{eq8}
          \end{eqnarray}
          
\begin{figure}[htb]
            \begin{center}
              \epsfig{file=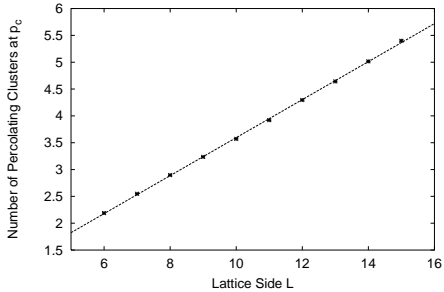,width=62mm}
            \end{center}
            \vspace{-1.5cm}
            \caption{\label{fig1}Random site percolation in seven dimensions: multiplicity of
          the spanning clusters at $p_c$ versus the linear dimension of the
          lattice.}
            \end{figure}

          In Fig. \ref{fig1} we plot the multiplicity of the percolating clusters
          as a function of the linear dimension of the lattice. The linear
          fit of the data is excellent. We conclude that,
          at least for the 
          particular case $d=7$, Eq. (6) is correct.

          \section{RESULTS FOR THE ISING MODEL}

          If there is a mismatch between the percolation and the thermal
          variables for $T<T_c$, the percolation transition of the FK clusters
          would show quite an unusual behaviour, since the critical
          exponents would
          be different above and below $T_c$.

\vspace*{-0.7cm}
          \begin{figure}[htb]
            \begin{center}
              \epsfig{file=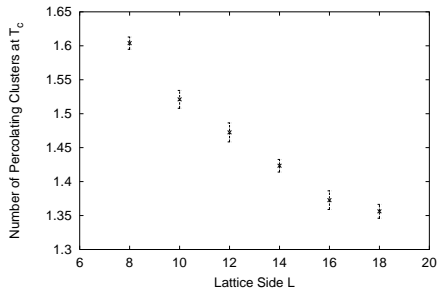,width=61mm}
            \end{center}
            \vspace{-1.5cm}
            \caption{\label{fig2}Ising model in five dimensions: multiplicity of the spanning
          FK clusters at $T_c$ versus the linear dimension of the lattice.}
          \end{figure}

\begin{figure}[htb]
            \begin{center}
              \epsfig{file=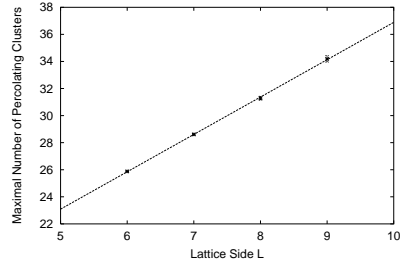,width=60mm}
            \end{center}
            \vspace{-1.5cm}
            \caption{\label{fig3}Ising model in seven dimensions: 
              scaling of the multiplicity peaks with the lattice side.}
            \end{figure}

\vspace*{-0.5cm}
          We investigated the Ising model in $5$, $6$ and $7$ dimensions. 
          We found that
          the behaviour below $T_c$ dominates, i.e. the finite-size scaling
          fits at $T_c$ return the values of the conjectured exponents for $T<T_c$. 
          Fig. \ref{fig2} shows the variation of the multiplicity $N_{\infty}$ of the percolating
          clusters with the lattice side $L$ for the 5-dimensional Ising model
          at $T_c$.
          We observe a clear decrease of $N_{\infty}$ with $L$, which
          is in contrast with Eq. (7)! On the other hand, since the
          scaling behaviour at $T<T_c$ dominates, this 
          result would confirm that the upper critical
          dimension $d^{FK}_u(T<T_c)>4$.
          We decided then to check whether $d^{FK}_u(T<T_c)=6$, 
          as conjectured. 
          For this purpose we investigated the 7-dimensional Ising model and
          we determined for each lattice
          the peak of the percolating cluster multiplicity. In Fig. \ref{fig3} we plot
          the values of the maxima  
          versus the lattice side $L$: the 
          points lie on a straight line, like in random percolation.

\vspace*{-0.5cm}

\vspace*{-0.8cm}
\section{CONCLUSIONS}

          Our results clearly show that  
          for random percolation hyperscaling breaks 
          due to the presence of infinite percolating 
          clusters at the critical point. 

          For the Ising model
          the situation is more complicated because of the 
          different behaviour of the percolation variables
          below and above $T_c$. We found that the 
          upper critical dimension of the FK clusters 
          below $T_c$ is six, like in random percolation.
          What really matters for our geometrical interpretation
          of hyperscaling breaking is a relation between
          $N_{\xi}$ and $\xi$ such that $N_{\xi}$ increases with $\xi$. 
          It is then crucial to extrapolate numerically such a relation 
          for $T\,{\neq}\,T_c$.
          Moreover, it is well known that the finite-size scaling behaviour 
          above the upper critical dimension is anomalous because
          the relevant scale is no longer the correlation length $\xi$ but
          another length $\ell$ \cite{luijten}. It would be interesting 
          to check whether $\ell$ coincides with the 
          above-mentioned geometrical fluctuation scale $\xi_1$.

%------------------------------------------------------------------------

\begin{thebibliography}{99}

\bibitem{coni} A. Coniglio, Physica A {\bf 281} (2000) 129.

\bibitem{FK} C. M. Fortuin, P. W. Kasteleyn,
            Physica {\bf 57} (1972) 536.

\bibitem{dearc} L. de Arcangelis, J. Phys. A {\bf 20} (1987) 3057.  

\bibitem{luijten} E. Luijten et al., Eur. Phys. J. B {\bf 9} (1999) 289.


\end{thebibliography}
\end{document}